\documentclass{ws-procs9x6}

\def\tr{{\rm Tr}\,} 
\def\L{{\rm L}} 
\def\R{{\rm R}}
\def\A{{\rm A}}
\def\c{{\rm color}}
\def\half{{\textstyle{1\over2}}}

\begin{document}

\title{Axions: Past, Present, and Future}

\author{Mark Srednicki}

\address{Department of Physics \\
University of California \\ 
Santa Barbara, CA 93106, USA}

\maketitle

\abstracts{I give a pedagogical and historical introduction to axion physics, 
and briefly review the present status of axions in our understanding of 
particle physics and cosmology.  This is a contribution to 
{\it Continuous Advances in QCD 2002/Arkadyfest}, held in honor of
Arkady Vainshtein's 60th birthday.}

\section{The Old and New U(1) Problems}\label{intro}

Consider quantum chromodynamics (QCD) with color group 
SU(3) and three flavors of massless quarks.   
The lagrangian is invariant under an additional global
flavor group, U$(3)_\L \times {}$U$(3)_\R$.  Although
we will use it anyway, this ``left--right''
terminology is actually somewhat misleading; in four dimensions,
a right-handed fermion field is the hermitian conjugate of a
left-handed fermion field, and so we could (and will)
adopt the convention that all fundamental fermion fields 
are left-handed (with, of course,
right-handed hermitian conjugates).  In this case, there would
be three left-handed fields transforming as $\bf 3$'s of 
SU$(3)_\c$,
and three left-handed fields transforming as $\bf \overline 3$'s of 
SU$(3)_\c$; thus the flavor group would be better named
U$(3)_{\bf 3} \times {}$U$(3)_{\bf\overline 3}$.  [We note in 
passing that if quarks were in a real or pseudo-real representation
of the color group, such as $\bf 3$ of O$(3)_\c$ or $\bf 2$
of SU$(2)_\c$, the lagrangian flavor group would be U(6),
since in this case there is no distinction between left-handed
quarks and left-handed antiquarks.]
For now we ignore the effects of the quantum anomaly, which will
of course play a key role later.

QCD exhibits dynamical breaking of the flavor symmetry; this
can be understood as the formation of a quark condensate,
\begin{equation}
\langle0| \chi_{\alpha i a}\tilde\chi^\beta{}_{\!\bar\jmath b}|0\rangle
= -{\textstyle{1\over6}}\Lambda^{\!3} \, \delta_\alpha{}^{\!\beta}
            \, \delta_{i \bar\jmath}
            \, \varepsilon_{ab}         \; .
\label{cond}
\end{equation}
Here $\chi_{\alpha i a}$
is a left-handed fermion field transforming as a $\bf 3$ of SU$(3)_\c$
with color index $\alpha=1,2,3$, ``left'' flavor index $i=1,2,3$,
and left-handed spinor index $a=1,2$; 
$\tilde\chi^\beta{}_{\!\bar\jmath b}$
is a left-handed fermion field transforming as a $\bf \overline 3$ 
of SU$(3)_\c$
with color index $\beta=1,2,3$, ``right'' flavor index $\bar\jmath=1,2,3$,
and left-handed spinor index $b=1,2$; 
$\varepsilon_{ab}$ is the antisymmetric invariant symbol of SU(2);
and $\Lambda$ is a parameter with dimensions of mass.
Under a U$(3)_\L$ transformation,
$\tilde \chi$ is unchanged, and
$\chi_{\alpha i a} \to L_i{}^j\chi_{\alpha j a}$,
where $L$ is a unitary matrix; 
under a U$(3)_\R$ transformation,
$\chi$ is unchanged, and
$\tilde\chi^\beta{}_{\!\bar\imath b} \to
(R^*)_{\bar\imath}{}^{\bar\jmath}
\tilde\chi^\beta{}_{\!\bar\jmath b}$,
where $R^*$ is an independent unitary matrix
(the complex conjugation is a notational convention).
The condensate is unchanged only by transformations in the
``vector'' subgroup U(3$)_{\rm V}$ specified by $R=L$.
Thus, U$(3)_\L \times {}$U$(3)_\R$ is spontaneously broken
down to U(3$)_{\rm V}$.
[I leave it as an exercise for the reader to show that the unbroken
subgroup would be O$(6)$ if quarks were
$\bf 3$'s of O$(3)_\c$, and Sp(6) if quarks were $\bf 2$'s
of SU$(2)_\c$.]  The nine broken generators lead us to expect
nine Goldstone bosons; these can be thought of as long wavelength
excitations of the condensate,
\begin{equation}
\langle0| \chi_{\alpha i a}(x)\tilde\chi^\beta{}_{b \bar\jmath}(x)|0\rangle
= -{\textstyle{1\over6}}\Lambda^{\!3} \, \delta_\alpha{}^{\!\beta}
            \, \varepsilon_{ab}     
            \, U(x)_{i \bar\jmath} \;,
\label{u}
\end{equation}
where $U$ is a unitary matrix field.  Under a
U$(3)_\L \times {}$U$(3)_\R$ transformation,
\begin{equation}
U(x)_{i \bar\jmath} \to
L_i{}^k \,
(R^*)_{\bar\jmath}{}^{\bar\ell} \,
U(x)_{k \bar\ell} \; ,
\label{url}
\end{equation}
or equivalently $U\to LU\!R^\dagger$.
We can write an effective lagrangian for $U$; it must of course be
U$(3)_\L \times {}$U$(3)_\R$ invariant.  There are no allowed terms
with no derivatives, and two with two derivatives:
\begin{equation}
L  = -{\textstyle{1\over4}}f_\pi^2 \,\tr\partial^\mu U^\dagger \partial_\mu U
            -{\textstyle{1\over18}}(f_1^2-{\textstyle{3\over2}}f_\pi^2)
                        \partial^\mu(\det U^\dagger)
                        \partial_\mu(\det U) +\ldots \; ,
\label{l}
\end{equation}
where $f_\pi$ and $f_1$ are parameters with dimensions of mass.
If we write 
\begin{equation}
U(x) = \exp\Bigl[ -i {\textstyle\sum_{a=1}^8} \lambda^a \pi^a(x)/f_\pi
                  -i \pi^9(x)/f_1 \Bigr] \;,
\label{u2}
\end{equation}
where the Gell-Mann $\lambda$ matrices are hermitian and normalized via 
$\tr\lambda^a\lambda^b = 2\delta^{ab}$,
then the hermitian Goldstone fields $\pi^a(x)$ 
($a=1,\ldots,9$) have canonical kinetic terms, 
$L = -\half \partial^\mu\pi^a\partial_\mu\pi^a$.
The parameter $f_\pi$ is called the {\it pion decay constant}, 
and it has the measured value of $92.4\,$MeV.  (It is measured
via the $\pi\to\mu\nu$ decay rate; see Ref.$\,$[\cite{georgi}]
for details.)

In the real world, the three light quarks have small masses:
\begin{equation}
L_{\rm mass} =  -M^{i\bar\jmath}\,\varepsilon^{ab}\,
                 \chi_{\alpha i a}
           \tilde\chi^\alpha{}_{b \bar\jmath} + {\rm h.c.} \;,
\label{mass}
\end{equation}
where $M$ is in general a complex matrix with no particular
symmetries; $M^\dagger M$ has eigenvalues $m_u^2$, $m_d^2$,
and $m_s^2$, where $m_{u,d,s}$ are the three light-quark masses.  
$M$ can be made diagonal with positive real
entries $m_u$, $m_d$, and $m_s$ via a
U$(3)_\L \times {}$U$(3)_\R$ transformation that leaves
the rest of the lagrangian unchanged.
In terms of the effective lagrangian, it is easy to see from 
eqs.$\;$(\ref{cond}) and (\ref{mass}) that
\begin{equation}
L_{\rm mass} = \Lambda^{\!3} \,\tr(MU + {\rm h.c.})\; .
\label{mass2}
\end{equation}
Expanding in the Goldstone fields, we get
\begin{equation}
L_{\rm mass} = -{\textstyle{1\over4}} \Lambda^{\!3} 
                \,\tr[M(\lambda^a\lambda^b+\lambda^b\lambda^a)]
                \pi^a\pi^b/f_\pi^2\; ,
\label{mass3}
\end{equation}
where $\lambda^9 \equiv (f_\pi/f_1)I$.
The key conclusion we want to obtain is reached most easily by
taking the exact isospin limit $m_u = m_d \equiv m$; then 
the eigenvalues of the mass-squared matrix in eq.$\;$(\ref{mass3}) are
$m_{1,2,3}^2 = 4\Lambda^{\!3}m/f_\pi^2$ (these are the three pions) and 
$m_{4,5,6,7}^2 = 2\Lambda^{\!3}(m+m_s)/f_\pi^2$ (these are the four kaons).
The $\pi^8$ and $\pi^9$ fields have a mass-squared matrix that must
be diagonalized; taking $m \ll m_s$, the eigenvalues are
$2\Lambda^{\!3}m_s(\frac43 f_\pi^{-2}+f_1^{-2})$
and $12\Lambda^{\!3}m / (f_\pi^2+\frac43 f_1^2)$.
This implies that there is a fourth pion-like particle with mass
less than $\sqrt3\,m_\pi$, where $m_\pi=m_{1,2,3}=135\,$MeV is
the neutral pion mass.  There is, in nature, no such particle.
This discrepancy between theory and experiment is the 
``U(1) problem''\cite{weinu1} or (nowadays) the ``old U(1) problem''.

The old U(1) problem is solved by the axial anomaly\cite{anomaly}: 
while the lagrangian is 
invariant under U$(3)_\L \times $U$(3)_\R$, the measure over the
fermion fields in the functional integral is not\cite{fuji}; the measure transforms
nontrivially under the ``axial'' U$(1)_\A$ subgroup $L=R^\dagger=e^{i\alpha}I$.
Furthermore, the presence of instanton solutions\cite{inst} of the euclidean
field equations of QCD allow us to see explicitly the physical effects of
the anomaly in semiclassical calculations\cite{semi}.
Thus, when we formulate an effective lagrangian for the low-energy
fields, we have no reason to expect it to be invariant under this
subgroup.  This allows us to add terms involving $\det U$ (and its derivatives)
to the lagrangian\cite{witten}:
\begin{equation}
L_{\rm anom} = f_\pi^4 \, {\textstyle\sum_{n=1}^\infty} 
               c_n\,(\det U)^n + {\rm h.c.} + \ldots\ .
\label{anom}
\end{equation}
In the absence of quark masses, the functional integral of QCD is CP invariant.
In the effective theory, CP exchanges $U$ and $U^\dagger$; CP invariance then
implies that the $c_n$ coefficients in eq.$\;$(\ref{anom}) must all be real.
The main effect of $L_{\rm anom}$ is to give the $\pi^9$ field a large mass,
$m_9^2 = 18Cf_\pi^4/f_1^2$, where $C=\sum_{n=1}^\infty c_n n^2$;
the $\pi^9$ field is then essentially removed from the effective theory,
and the mass of the $\pi^8$ field (physically, the $\eta$ meson) is given by
${\frac83}\Lambda^{\!3}m_s/f_\pi^2$.

A shadow of the $\pi^9$ field remains, however; 
in the presence of $L_{\rm anom}$, we cannot
remove the overall phase of the quark mass matrix $M$ via a    
U$(3)_\L \times $U$(3)_\R$ transformation.  
If $\det M = m_u m_d m_s e^{i\theta}$, and if we then make a
U$(3)_\L \times $U$(3)_\R$ transformation on $U$ that brings $M$ to the form
${\rm diag}(m_u,m_d,m_s)$, the $c_n$'s change: $c_n\to c_n e^{-in\theta}$.
This would result in CP violating effects in hadron interactions;
the most visible of these would be an 
electric dipole moment for the neutron\cite{demon}.
Experimental limits on the electric dipole moment of the neutron imply
$\theta < 10^{-9}$.  This, then, is the ``strong CP problem'', or the 
``new U(1) problem'': why is $\theta$ so small?

One possibility is that $\det M=0$ because $m_u=0$; then there is no phase
to remove from the quark mass matrix.  It turns out that this possibility 
cannot be ruled out solely by experimental evidence (since higher-order
corrections in the $d$ and $s$ quark masses can mimic a nonzero $u$ quark
mass\cite{km}), but it does require some severe theoretical contortions
(such as the invalidity of the large-$N_{\rm color}$ expansion to all orders);
see Ref.$\,$[\cite{leut}] for a discussion.

\section{The Peccei--Quinn Solution}\label{pq}

Let us choose to work in a field basis in which the $c_n$'s are real 
and the quark-mass matrix $M$ is diagonal, but contains the unremovable 
phase $\theta$:
$M={\rm diag}(m_u e^{i\theta}, m_d, m_s)$.
Now the potential energy is minimized at nonzero values of 
$\pi^3$, $\pi^8$, and $\pi^9$.
For small $\theta$, this minimum energy works out to be
\begin{equation}
V_{\rm min}(\theta) = {m_u m_d m_s \over m_u m_d + m_d m_s + m_s m_u } \,
                      \Lambda^{\! 3} \, \theta^2 \;.
\label{min}
\end{equation}
This suggests a possible resolution of the strong-CP problem.
If $\theta$ was some sort of dynamical variable,
then it would want to relax to zero, in order to minimize the energy.
This is the essential physics of the solution proposed by 
Peccei and Quinn\cite{pq}.  In their model, $\theta$ is effectively
replaced by $a(x)/f_a$, where 
$a(x)$ is a new scalar field, the {\it axion}, and $f_a$ is a new parameter, 
the {\it axion decay constant}.  The observed
value of $\theta$ is now zero, since that is what minimizes the energy.

Of course, this new field provides us with a new particle as well.
Its mass is given by 
\begin{eqnarray}
m_a^2 &=& {1\over f_a^2} {d^2 V_{\rm min}(\theta)\over 
                          d\theta^2}\biggr|_{\theta=0} 
\nonumber \\
&=& {2 m_u m_d m_s \over m_u m_d + m_d m_s + m_s m_u } \,
{\Lambda^{\! 3}\over f_a^2} 
\nonumber \\
&=& \left[{ m_u m_d m_s / (m_u+m_d)
           \over m_u m_d + m_d m_s + m_s m_u }\right]
    {f_\pi^2\over f_a^2} \, m_\pi^2 \;,
\label{amass} 
\end{eqnarray}
where we have used $m_\pi^2 = 2(m_u+m_d)\Lambda^{\!3}/f_\pi^2$ 
to get the last line.  
The factor in brackets has the numerical value\cite{leut} $0.225\pm0.005$.
We see that if $f_a \gg f_\pi$, then the axion is very light.

Peccei and Quinn did not present their solution in this language.
They were concerned with constructing a renormalizable extension
of the Standard Model of quarks and leptons
that did not suffer from the strong CP problem.  
In their model, they used two Higgs fields, 
instead of one: one to give mass to the up quarks, 
and the other to give mass 
to the down quarks (and, somewhat incidentally, the charged leptons).  
The axion then appears as the relative phase of these two
fields, and it is massless (classically) if the lagrangian
is invariant under an extra U(1) symmetry,
now known as {\it Peccei-Quinn symmetry},
that is spontaneously broken by the Higgs vacuum expectation values.
In working out the physics, Peccei and Quinn assumed
that the anomaly would give a large mass to the axion; the fact
that it has a small mass was noticed by Weinberg\cite{wein} 
and Wilczek\cite{wilc}.
In the Peccei-Quinn model, the axion decay constant is given by 
$f_a = {1\over6}\sin(2\beta)v_{\rm EW}$,
where $v_{\rm EW}=(\sqrt2 G_{\rm F})^{-1/2}=246\,$GeV is the electroweak scale,
and $\tan\beta$ is the ratio of the two Higgs-field VEVs.
The couplings of the axion to quarks, leptons, and photons are
all proportional to $1/f_a$, and hence of typical weak-interaction
strength.  Thus the axion should be roughly as visible as a
neutrino.  It soon
became apparent that there is no such particle in nature.

\section{Invisible Axions}\label{invis}

From our presentation in the last section, it should be clear
that there is no fundamental principle that relates $f_a$ to
$v_{\rm EW}$; this is simply a consequence of the specific model
explored by Peccei and Quinn.  Making $f_a$ larger would make
the axion lighter, but also more weakly coupled to ordinary
matter of all kinds, and hence harder to produce or detect.  


The first successful theoretical proposal to detach $f_a$ from 
$v_{\rm EW}$ came from Kim\cite{kim}.  
His model is just the Standard Model 
(with the usual one Higgs doublet), 
plus an additional sector consisting
of a Higgs field that is an SU(3)$\times$SU(2)$\times$U(1) singlet,
with a Yukawa coupling to a new quark field that is an SU(2) singlet and carries
an essentially arbitrary hypercharge.  (Thus, the electric charge
of this quark is equal to its hypercharge.)  The Peccei-Quinn symmetry
of this model involves only these new fields.
The singlet Higgs field
gets a vacuum expectation value $v$, breaking the PQ symmetry.
Since this Higgs VEV does not break any gauge symmetries, it can
be arbitrarily large.  If $v \gg v_{\rm EW}$, then the
phase of this field becomes the axion, and $f_a \simeq v$.
The new quark gets a mass given by $v$ times the value of the 
Yukawa coupling.  
Kim's paper was submitted to {\it Physical Review Letters}
in February 1979 and published in July 1979.

Next came the proposal of Zhitnitski\u\i\cite{zhit};
his model consisted of the Peccei-Quinn model (with its two
Higgs doublets), plus an extra singlet Higgs
field with (as in Kim's model) a large VEV $v$.  This Higgs singlet
is coupled to the two doublets in such a way as to preserve the
Peccei-Quinn symmetry of the original lagrangian.
The large singlet VEV spontaneously breaks the PQ symmetry,
and (as in Kim's model) this results in the phase of this
field becoming the axion, with $f_a \simeq v$.
Zhitnitski\u\i's paper was submitted to {\it Yadernaya Fizika}
in May 1979 and published in February 1980.  The
English translation (by Jonathan Rosner) appeared
in the {\it Soviet Journal of Nuclear Physics} about a year later.

Neither of these papers attracted much attention at the time.
I will now play amateur sociologist and attempt to guess why.

The title of Kim's paper,
``Weak-Interaction Singlet and Strong CP Invariance,''
 gives little clue as to its content.
The abstract, however, is clear and succinct:
``Strong CP invariance is {\it automatically\/} preserved by
a spontaneously broken chiral U(1$)_\A$ symmetry.
A weak-interaction singlet heavy quark $Q$, a new scalar
meson $\sigma^0$, and {\it a very light axion} are predicted.
Phenomenological implications are included.''
In the introduction, Kim takes a different tack,
beginning with a classification scheme for 
theories of CP violation (his categories are {\it hard},
{\it dynamical}, and {\it spontaneous}), and remarking that
``a {\it simple} theory should belong to one of these classifications
for {\it completeness} of the gauge theory of weak and
electromagnetic interactions.''  
He then presents his model.
However, Kim does not emphasize the arbitrariness of $f_a$;
when he needs a numerical value, he uses $f_a=100\,$TeV,  
and he takes 100$\,$GeV for the mass of the new quark.  
He considers the new quark (and the heavy radial excitation
of the new Higgs, $\sigma^0$) to be important aspects of
the model's phenomenology.  He concludes:
``What are the possible experiments to prove the present
scheme?  Probably high-precision experiments of the axion
search will do.  But the easier verification of the
weak-interaction singlets $Q$ and $\sigma^0$ 
in $pp$ or $\bar p p$ machines ($e^+ e^-$ annihilation
machine also if $Q$ is charged) will shed light on the
whole idea of the spontaneously broken chiral
U(1$)_{\rm A}$ invariance and the multiple vacuum
structure of QCD.''  A new quark at the electroweak scale
with nonstandard electroweak interactions that could not be
fit into grand unification schemes must have seemed to many 
at the time to be an unattractive proposition. 

Zhitnitski\u\i's paper, titled ``On the possible
suppression of axion-hadron interactions,''
is clear, succinct, and correct throughout.  His abstract
reads, ``A possible mechanism for strong suppression
of the axion-fermion interaction is considered.
Two models in which this mechanism is realized
are described in detail.''  The emphasis is on
the axion coupling to quarks; the couplings to leptons
and photons (which are also suppressed) are not
mentioned.  Also, the axion mass is mentioned only 
in the paper's last sentence:
``Here the mass of the axion is
$m_a \sim 100\,\delta\;$keV,''
where $\delta\simeq v_{\rm EW}/v$.
In addition to the model discussed above,
Zhitnitski\u\i~also presents a model based on the gauge
group SU(3)$\times$SU(2$)_{\rm L}\times$SU(2$)_{\rm R}\times$U(1);
in this model, no new Higgs fields are needed other
than the ones already needed to break SU(2$)_{\rm R}$.
The emphasis on a left--right symmetric model
probably contributed to this paper's initially limited impact;
these models were by then waning in popularity.

Next came the paper entitled
``Can confinement ensure natural CP invariance of the strong interactions?''
by Shifman, Vainshtein, and Zakharov (SVZ)\cite{svz}, which was submitted to
{\it Nuclear Physics B\/} in July 1979 and published in April 1980.  
SVZ point out that the
usual argument for the physical effect of the $\theta$ parameter rests on the
properties of long-range instanton field configurations, and it is not immediately
obvious how to treat these in the presence of color confinement.
Could it be that color confinement renders $\theta$ unobservable after all?
Their answer is {\it no}; my presentation in Sect.$\,$1 of low-energy physics with
$\theta\ne0$ essentially follows their argument.
They show, in particular,
that nonzero $\theta$ results in a nonzero amplitude for $\eta\to\pi\pi$   
decay.  So the effect of $\theta$ is really there, and the strong CP problem is
really a problem.  This analysis makes up the bulk of the paper.

So, what to do?  SVZ begin the last section of their paper by stating that,
``In fact, the problem of the $\theta$-term cannot be solved by QCD alone
since it is intimately related to the origin of the quark masses and the
mass generation is considered to be a prerogative of the weak interactions.''
They then note than even the phase of an arbitrarily heavy quark would
contribute to $\theta$ (which is just the phase of the determinant of the 
quark mass matrix),
and that this would even apply to the mass of a Pauli-Villars regulator field!
``Since $\theta$ is directly observable (see above),'' they write,
``it seems to be a unique example of how a laboratory experiment can
shed light on regulator properties.  Alternatively, it is very suspicious.''

Their suspicions aroused, they then proceed to construct a model 
identical to Kim's.  They conclude their paper with this paragraph:
``If $v$ becomes arbitrarily large, the axion interaction with normal
hadrons vanishes.  (The same is true for its mass.)  The new quark also
becomes unobservable.  This theoretical phantom still restores the natural
P and T invariance of QCD.  Although the model discussed is evidently
a toy one, one might hope that something of the kind happens in 
unified theories.''

Here, then, is a clear presentation of the key point, by well known
and respected physicists, published in the most widely read particle-physics
journal of the time.  Still, the initial impact was minimal.  Why?

One clue can be found in this paragraph, near the top of the paper's second page:
``The overwhelming reaction to this observation is that the example of the one-instanton
solution does prove CP violation in the presence of the $\theta$-term.  Our final
result complies with the conclusion on CP violation so that most readers can,
justifiably, lose interest in the paper at this point.''
Presumably, most readers did just that!

The fourth discovery of the arbitrariness of $f_a$ came in the paper
``A simple solution to the strong CP problem with a harmless axion,''
by Dine, Fischler, and myself\cite{dfs}, submitted to {\it Physics Letters B}
in May 1981 and published in August 1981.  We had been working on 
(very messy) models of supersymmetric technicolor, and
had found an axion with a large decay constant in one of them.
We began wondering whether this phenomenon could be duplicated
in a simple Higgs model, and quickly constructed one
(which turned out to be identical to Zhitnitski\u\i's).  
We were excited at first, then wary.  This was too easy!  
If it really worked, surely others would have thought of it before.

So we began trying to search the literature (without, of course,
a computer database).  We found Kim's paper pretty quickly.
We were disappointed, but luckily our specific model was different:
we didn't have an extra quark, only extra Higgs fields.
We felt this would be considered more palatable.
Also, we could fit our model into the GUT paradigm of the day.
Our abstract reflects these points: ``We describe a simple generalization
of the Peccei-Quinn mechanism which eliminates the strong CP problem 
at the cost of a very light, very weakly coupled axion. 
The mechanism requires no new fermions and is easily implemented
in grand unified theories.'' 
As the paper was being typed (by a {\it secretary}, on a {\it typewriter}),
we continued to poke around in
the library, because we were still worried.  Eventually we came
across the SVZ paper.  We were relieved to find that their model
was the same as Kim's.  In our text, we talked about ``Kim's model,''
and obviously we needed to change this to ``the model of Kim and
Shifman, Vainshtein, and Zakharov''.
Too late!  Our paper was typed, and re-typing was out of the question.
We were at least able to add ``see also SVZ'' at the end of the reference
to Kim.  It came out looking as though we thought the SVZ paper
was somehow inferior to the Kim paper.  I don't know if anyone ever
noticed this, but I would like to take this occasion 
of Arkady's sixtieth birthday to apologize to him and his collaborators!

Much to our surprise, our paper received a great deal of attention.
Wise, Georgi, and Glashow\cite{wgg} constructed an explicit SU(5) model,
and made the important point that solving the strong CP problem
in this way did not require any extra fine-tuning in the Higgs sector.
They also coined the phrase ``invisible axion,'' which has stuck.
Nilles and Raby constructed a similar supersymmetric model\cite{nr}.
Many more papers followed.
The two original models became known as the KSVZ and DFS models.
Zhitnitski\u\i's paper was still generally unknown and unreferenced.

In 1986, I had a (wrong, as it turned out) idea about the electric 
dipole moment of the neutron, and consulted Shalabin's 1983 
review\cite{shal}.  In it was a reference to Zhitnitski\u\i, 
right where I expected a reference to DFS.  Wondering what
this paper could be, I looked it up,
began reading, and felt a flush of embarrassment wash over me as
I realized that (a) it had a model identical to ours and (b) it
was written and published long before ours.
I began to reference it (and to ask others to reference
it in axion papers that I refereed).  I don't know if this was
the catalyst or not, but eventually the DFS model became the DFSZ model.

\section{Axion Astrophysics and Cosmology}\label{cosmo}

Detailed discussions of the axion couplings to hadrons, electrons,
and photons can be found in Ref.$\,$[\cite{agg}].  All are proportional
to $1/f_a$, and the strongest limits on them come from astrophysical
processes: axion emission from various objects (red giants,
stars in globular clusters, supernova 1987A, etc.) must be slow 
enough to avoid changing the physics.
The results are model dependent, but roughly require $f_a > 6\times10^9\,$GeV,
or $m_a < 0.01\,$eV.  For a review, see Ref.$\,$[\cite{axas}].

The axion can play a cosmological role\cite{axcos} for larger values of $f_a$.
At high temperatures, the instanton effects which give rise to the
axion mass go away.  In the very early universe, then, the value
of $\theta$ is undetermined, and presumably varies slowly from
place to place.  Inflation would pick out a particular value in our
horizon volume.  If subsequent reheating is to a temperature below
$f_a$, this value is frozen in place until much later, when the
universe cools and the axion mass appears.  The axion field then
begins coherent oscillations around its minimum.  If the initial
value of $\theta$ is $O(1)$, then the energy stored in these oscillations
would overclose the universe if $f_a$ is larger than about $10^{12}\,$GeV.
If $f_a$ is near this value, this energy (which is in the form of 
axions with near zero momentum) could be the cold dark matter.

Things are more complicated if there is no inflation, or reheating
after inflation to a temperature above $f_a$.  In this case a network
of cosmic-string defects forms in the axion field\cite{davis}, and the evolution
of this network is a complicated numerical problem.  The axions from
string decay contribute at least as much dark matter as the axions
from the initial misalignment, and possibly much more; see
Ref.$\,$[\cite{axas}] for more details.

The possibility that axions form the cold dark matter leads to the
exciting possibility that we might be able to detect them.
The axion-photon interaction lagrangian is
\begin{equation}
L_{a\gamma\gamma} = {\alpha\over 2\pi f_a}
                    (C-C')a{\bf E\!\cdot\!B} \;,
\label{agg}
\end{equation}
where $C$ and $C'$ are numerical constants.
This interaction allows an axion to convert to a photon in a magnetic
field, and Sikivie proposed using this effect to search for
dark-matter axions\cite{sikivie}.
The value of $C$ is 
computed in terms of the Peccei-Quinn charges
of the fermions, and $C'$ arises from
axion-pion and axion-eta mixing; it is model independent, and given by
\begin{eqnarray}
C' &=& {2\over3}\,{m_u m_d + 4 m_d m_s + m_s m_u \over
                   m_u m_d +   m_d m_s + m_s m_u} 
\nonumber \\
&=& 1.93 \pm 0.04\; .
\label{cmix}
\end{eqnarray}
Unfortunately, $C$ is typically positive, so there
is a cancelation.  In grand-unifiable models (that is, models
where the fermions come in SU(5) multiplets with the same
PQ charge; whether the extra heavy particles implied by
unification exist or not is irrelevant) such as the DFSZ model,
\begin{equation}
C_{\rm DFSZ}  = {8\over 3} \;.
\label{cpq}
\end{equation}
In KSVZ models with a single heavy quark of charge $Q$,
\begin{equation}
C_{\rm KSVZ}  = 6Q^2 \;.
\label{cpq2}
\end{equation}
If axions are the dark matter, $m_a \sim 1$ to $100\,\mu$eV is the most
interesting mass range ($1\,\mu{\rm eV} = 10^{-6}\,$eV). 
Experiments to search for them are currently underway,
and one\cite{hag} has ruled out KSVZ axions (with $Q=0$) in the mass range
$2.8 \pm 0.5\;\mu$eV as a significant component of the local dark matter.  
The experiments are continuing; for a review, see Ref.$\,$[\cite{axexp}].

\section{The Future of Axion Physics}\label{future}
Over twenty years have passed since the invention of the invisible axion,
and we still do not know whether or not this is the correct solution
to the strong CP problem.  It is vitally important that the current
searches continue until they have covered at least the most plausible mass
and coupling ranges; it would be a great shame if such important physics
surrounded us, and we left it undiscovered.  Of course, even if these searches
do not find dark matter axions, this means only that the dark matter is not 
axions, and not that axions do not exist.  There are many cosmological
scenarios for this possibility.

Also, it should be noted that axions arise rather naturally in superstring
models, and it may be that any ground state of string/M theory that resembles
the Standard Model always includes an axion with, say, 
$f_a \sim (v_{\rm EW}M_{\rm Planck})^{1/2} \sim 10^{11}\,$GeV.
However at present we seem a long way from being to reach this sort of conclusion.

The road ahead for axion physics is thus likely to be a hard one, both theoretically and
experimentally.  But the reward for a successful traversal will make the
journey worthwhile.

I would like to thank the organizing committee for Arkadyfest---Keith Olive,
Mikhail Shifman, and Mikhail Voloshin---for the opportunity to present this
talk, and Arkady Vainshtein for providing the occasion and the inspirational
physics.  I also acknowledge the NSF for financial support.

\end{document}